\documentclass[journal]{IEEEtran}
\usepackage{array}
\newcolumntype{P}[1]{>{\centering\arraybackslash}p{#1}}

\ifCLASSINFOpdf
  \usepackage[pdftex]{graphicx}
\else
\fi
\usepackage[cmex10]{amsmath}
\newcommand{\qed}{\nobreak \ifvmode \relax \else
      \ifdim\lastskip<1.5em \hskip-\lastskip
      \hskip1.5em plus0em minus0.5em \fi \nobreak
      \vrule height0.75em width0.5em depth0.25em\fi}

\usepackage{subcaption}
\usepackage{kantlipsum} %<- For dummy text
\usepackage{mwe} %<- For dummy images

\DeclareCaptionLabelSeparator{periodspace}{.\quad}
\usepackage{mdwmath}
\usepackage{mdwtab}
\usepackage{caption}
\begin{document}
\title{JPEG Image Compression using the Discrete Cosine Transform: An Overview, Applications, and Hardware Implementation}
\author{\IEEEauthorblockN{
Ahmad~Shawahna,\IEEEmembership{}
Md. Enamul Haque,\IEEEmembership{} and
Alaaeldin Amin\IEEEmembership{}\\
}
Department of Computer Engineering\\
King Fahd University of Petroleum and Minerals, Dhahran-31261, KSA\\ 

\{g201206920, g201204920, amindin\}@kfupm.edu.sa
}

\maketitle

% =========================ABSTRACT=========================

\begin{abstract}

Digital images are becoming large in size containing more information day by day to represent the as is state of the original one due to the availability of high resolution digital cameras, smartphones, and medical tests images. Therefore, we need to come up with some technique to convert these images into smaller size without loosing much information from the actual. There are both lossy and lossless image compression format available and JPEG is one of the popular lossy compression among them. In this paper, we present the architecture and implementation of JPEG compression using VHDL (VHSIC Hardware Description Language) and compare the performance with some contemporary implementation. JPEG compression takes place in five steps with color space conversion, down sampling, discrete cosine transformation (DCT), quantization, and entropy encoding. The five steps cover for the compression purpose only. Additionally, we implement the reverse order in VHDL to get the original image back. We use optimized matrix multiplication and quantization for DCT to achieve better performance. Our experimental results show that significant amount of compression ratio has been achieved with very little change in the images, which is barely noticeable to human eye.
\end{abstract}



% \fi
\IEEEpeerreviewmaketitle

%==============================INTRODUCTION=========================================================================================================================================================

\section{Introduction}

\IEEEPARstart{J}{PEG} stands for Joint Photographic Expert Group. This is the most popular and common form of image compression. If we think about how the images are represented in terms of numbers, we would find out that every pixel values are represented as numbers. Those values have certain intensity values to represent them as red, green or blue for color images. The concept is same for black and white images as well, except there exists only two types of numbers (zero and one). Those images which have higher number of pixels turn out as good quality image. The intensity map of the matrix for the pixel value of an image is constructed with the bit depth. As, the technology is growing rapidly, digital images are becoming large with more pixels.   

Image compression technique is achieved by using statistical inference from the image pixel values. There exists significant redundancies of pixel values in each observable image. Human eye can not differentiate much if some of the information from the original image becomes absent. From this phenomena, image compression technique evolved and playing significant role in minimizing cost and bandwidth in digital arena. JPEG compression takes place in five steps with color space conversion, down sampling, discrete cosine transformation (DCT), quantization, and entropy encoding. DCT transformation is used due to its energy compaction characteristics. DCT has cosine function which is easier to compute and the number of coefficients become less. Thus, DCT can result more accurate image reconstruction even if the JPEG is lossy transformation. There is one step called quantization where less important pixels are discarded according to the frequency distribution. This remaining pixels form the compressed image. So, some distortion is generated afterwords, but the level of distortion can be adjusted during the compressing from quantization matrix [7]. We have chosen to use 1-D DCT on both row and columns to make 2-D DCT effective. There are some common quantization matrix available that can be used to vary the output. 

Discrete cosine transform is the fundamental part of JPEG [6] compressor and one of the most widely used conversion technique in digital signal processing (DSP) and image compression. Due to the importance of the discrete cosine transform in JPEG standard, an algorithm is proposed that is in parallel structure thus intensify hardware implementation speed of discrete cosine transform and JPEG compression procedure. The proposed method is implemented by utilizing VHSIC hardware description language (VHDL) in structural format and follows optimal programming tips by which, low hardware resource utilization, low latency, high throughput and high clock rate are achieved.

The motivation behind DCT image compression is that JPEG compression has become one of the most popular techniques for image compression and is being used in a wide variety of applications. It is involved in digital cameras, the digital altering of images, loading pictures on the web and various other applications. Nowadays, the focus has shifted to using reconfigurable hardware to implement the JPEG algorithm to increase its efficiency and hence reduce the cost of this technique. Significant amount of research is going on in this area and our aim through this paper is to achieve JPEG compression using VHDL with better performance. This method can be very useful in medical image storage or traffic image storage as they need colossal amount of images to be stored daily. So, fast and reversible compression method can be very useful in these areas. Our aim is to implement the DCT for JPEG images and evaluate the performance for different quantization matrix and compare the result with other related research.

The remainder of this paper is organized as follows. In section II, we describe the standard DCT based JPEG compression and decompression method. Section III represents concise overview of the related work done in this area. In section IV we present the implementation details of our work. Section V, VI and VII focuses on the controller, data path, simulation and synthesis result respectively. Finally, we conclude and suggest future direction of our work in section VII.

%==============================SECTION 2 =========================================================================================================================================================

\section{Overview Of Digital Image Compression and Decompression}

When we think about image compression it is apparent that the total number of bits present in an image can be minimized by removing the redundant bits. There are three types of redundancy available in terms of space, time and spectrum. Spatial redundancy indicates the correlation between neighboring pixel values. Spectral redundancy indicates correlation among different color planes. Temporal redundancy indicates correlation among different frames in an image. Compression techniques or methods aim to reduce the spatial and spectral redundancy with maximum efficiency. 

The compressed image quality depends on the compression ratio [19] of the original and compressed image. The compression ratio $(CR)$ is defined as,

\begin{equation}
  \begin{split}
  CR
    &= \frac{n_1}{n_2}
     \end{split}
\end{equation}

Where, $n_1$ and $n_2$ refers to number of information carrying units in original image and compressed image respectively. 

Relative data redundancy, $RD$ of the original image shows three possibilities of the pixel redundancy. The relationship with compression ratio is defined as,

\begin{equation}
  \begin{split}
  RD
    &= 1 - \frac{1}{CR}
     \end{split}
\end{equation}
Both relative data redundancy and compression ratio can express the below possibilities,

\begin{enumerate}
\item When $n_1 = n_2$, then $CR=1$ and hence $RD=0$. Thus no redundancy in the original image. 
\item When $n_1 \gg n_2$ ,then $CR \to\infty$ and hence $RD > 1$. Thus there is sufficient redundancy in the original image.
\item When $n_1 \ll n_2$ ,then $CR > 0$ and hence $RD \to \infty$. Thus the compressed image contains more data than original image.
\end{enumerate}

Digital image compression turns the pixels less correlated than they were before. The compression and decompression techniques are opposite of each other. There are different algorithms and techniques for digital image transformation, e.g, Discrete Fourier Transformation (DFT), Fast Fourier Transformation (FFT), Wavelet Transformation, Fractal compression etc. We have chosen discrete cosine transformation (DCT) [18] for the improved performance and less complexity.

\subsection{Compression}
\begin{figure}[!h]
  \centering           
        \includegraphics[width=0.48\textwidth]{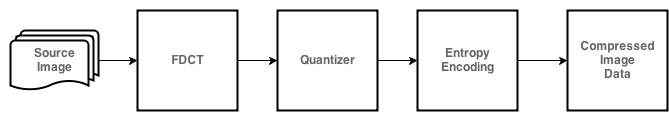}   
        \caption{Image compression steps.} 
        \label{comp}
\end{figure}

The JPEG compression process is broken into three primary parts as shown in Figure~\ref{comp}. To prepare for processing, the matrix representing the image is broken up into 8x8 squares and passed through the encoding process in chunks.  Color images are separated into three different channels (each equivalent to a greyscale channel) and treated individually.

The input matrix from the images each pixel values are subtracted by 128 before the quantization using the below equation.

\begin{equation}
  \begin{split}
  P(\mathbf{x,y})
    &= P_{i,j} - 128
     \end{split}
\end{equation}

Where, $i$ and $j$ are the row and column numbers specific to the pixel values. For example, $i=1024$ and $j=768$ for an $1024 * 768$ pixel image.

The 2-D DCT is implemented using row-column decomposition technique. Initially, 1-D DCT for each column and later for each row is computed. Input data matrix is multiplied with the coefficient matrix and then the result is multiplied with the transpose of the coefficient matrix. 

Next, the values of the matrix are quantized using some common quantization matrix. Quantization is the major part of the compression steps as it reduces the quality of the image by scaling down the original values. So, if the desired compression needs better quality, the tweaking should be in the quantization matrix. General formula for using the quantization matrix according to the scaling is given below.

 \begin{equation}
Q_{50} = \begin{bmatrix}
       16 &  11 &  10 &  16 & 24 & 40 & 51 & 61 \\
       12 &  12 &  14 &  19 & 26 & 58 & 60 & 55 \\
       14 &  13 &  16 &  24 & 40 & 57 & 69 & 56 \\
       14 &  17 &  22 &  29 & 51 & 87 & 80 & 62 \\
       18 &  22 &  37 &  56 & 68 & 109 & 103 & 77 \\
       24 &  35 &  55 &  64 & 81 & 104 & 113 & 92 \\
       49 &  64 &  78 &  87 & 103 & 121 & 120 & 101 \\
       72 &  92 &  95 &  98 & 112 & 100 & 103 & 99\\
    \end{bmatrix}
\end{equation}

The quality of the reconstructed image is adjusted by varying this matrix. Typically the quantized matrix which is obtained from quantization has values primarily in the upper left (low frequency) corner. By using a zigzag ordering to group the non zero entries and run length encoding, the quantized matrix can be much more efficiently stored compared to the non-quantized version.

% 2.2 DECOMPRESSION SUBSECTION
%==============================
\subsection{Decompression}

\begin{figure}[!h]
  \centering           
        \includegraphics[width=0.48\textwidth]{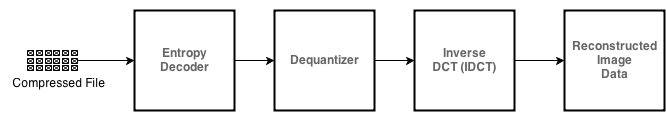}   
 \caption{Image reconstruction (Decompression) steps.}
 \label{Decompressionimg}
\end{figure}

Decompression is the reverse process of compression. At first, the decoder takes the compressed image data as its input. It then applies a run length decoding, inverse zigzag, de-quantization, inverse discrete cosine transform (IDCT), then obtains the reconstructed image. Figure~\ref{Decompressionimg} shows the steps for image reconstruction (decompression).

Run length decoding will have to perform the inverse process of run length encoding. Run length decoding produces original data stream as output. This linear data stream is converted into matrix format using inverse zig-zag ordering for every 8X8 block. Next, inverse quantization is done by multiplying the standard quantization matrix, $Q(u,v)$ with resultant quantized value to get the inverse dot value for each 8X8 matrix.

\begin{equation}
IDCT(\mathbf{u,v}) = Q(u,v) Q_{DCT}
\end{equation}

After computing the $IDCT$, the signed output samples are level-shifted. This level shifting converts the output to an unsigned representation. For 8-bit precision, the level shift is performed by adding 128 to every element of the block from the $IDCT$ output.

\begin{equation}
P(\mathbf{x,y}) = round \Bigg(\frac {IDCT(\mathbf{u,v})}{CC^T} + 128 \Bigg)
\end{equation}

It is very usual that the decompression process may produce values outside of the original input range of [0, 255]. When this occurs, the decoder needs to trim the output values to keep within that range. The decompressed image can be compared to the original image by taking the difference.

% RELATED WORK
%============================%============================
\section{Related Work}

\textbf{Patidar} et al. [1] removed the redundancy in the image data using 2-D DCT through two 1-D DCT performed on 8X8 matrix. They converted the source image into minimum code units and applied this 2-D DCT on each block. Then they used quantization technique to reduce the number of discrete symbols in a given stream to make the image more comprisable. They performed zig-zag reordering after the quantization stage to similarize with the transpose buffer used. Finally both Huffman and run-length encoding is performed based on statistical characteristics.

\textbf{Deepthi} et al. [2] worked on both compression and decompression of JPEG images. Their flow constituted with image segmentation and downsampling, DCT transformation, quantization and encoding for compression. Image segmentation and downsampling was done after loading the images each block of 8X8 pixels as minimum code unit. The authors main interest was the hardware implementation of the 2-D DCT combined with quantization and zig-zag process. They used pipelined architecture rather than single clock architecture to achieve high throughput.

\textbf{Kumar} et al. [3] described the design of two-dimensional discrete cosine transform (DCT) architecture for Multimedia communication applications. Their transformation method transforms images from spatial domain to frequency domain. Their main objective was to explore available architectures for optimizing the area and performance. They designed one architecture and implemented that in VHDL and synthesized using Xilinx tools and implemented on FPGA.

\textbf{Enas} et al. [4] used the architecture of 2-D DCT with quantization and zig-zag arrangement to compress images in VHDL. This architecture calculated the DCT using scaled DCT. The real DCT coefficient was obtained by multiplying the post-scalar value and DCT module. 2-D DCT was computed by combing two 1-D DCT. Their design used 3174 gates, 1145 slices, 21 I/O pins and 11 multipliers of one Xilinx Spartan-3E FPGA with operating frequency 84.81 MHz. One input block with 8x8 elements of 8 bits each is processed in 2470 ns and pipelined latency is 123 clock cycles.

\textbf{Frid} et al [9] mentioned about the alternative ways of implementing the DCT transformation algorithm in software. They compared the result of and IEEE standard 1180 definition and results from FPGA development boards such as Spartan-3E and Virtex-5 with 32-bit MicroBlazeTM soft-core processor. They presented AAN algorithm implementation on software, a special FPGA ip-core that accelerates the standard DCT algorithm and AAN algorithm (One dimensional post scaled DCT algorithm).

\textbf{Mankar} et al. [11] proposed one 1-D DCT using NEDA (New Distributed Arithmetic) for implementing inner products without using multipliers and ROM for converting signal into elementary frequency components. They accumulated the outputs at every clock cycles. They obtained maximum frequency 311.943 MHz and throughput 3431.384 Mbit/s. Their proposed design consumes 2.76 W at its maximum frequency.

\textbf{Anitha} et al. [12] proposed a novel algorithm for computing 2-D FFT and inverse FFT for realizing on hardwares. They emphasized mostly on processing speed. They used MATLAB to code both FFT and IFFT algorithms for 2D color images. The reconstructed images were identical with the source images and the quality is better than 35dB.

\textbf{Yamatani} et al. [13] proposed two new image compression-decompression methods for producing better visual accuracy, PSNR for low bit rates. The first method, the "full mode" \textit{polyharmonic local cosine transform} (PHLCT) modifies the encoder and decoder of the baseline JPEG method. The aim of the first method was to reduce the code size in encoding and blocking artifacts for decoding part. The second method, "partial mode" PHLCT modifies only the decoder part.

\textbf{Trang} et al. [14] proposed a high accuracy and high-speed 2D 8X8 Discrete Cosine Transform design. They used parallel matrix multiplication for 8X8 pixel blocks. They also mentioned that the implementation is not as accurate as standard references like MATLAB (after rounding) and Xdiv MPEF 4. Their proposed design was implemented on Xilinx Virtex 4 and could run at 308 MHz. The image processing rate was 145 fps (frames per second).

\textbf{Bukhari} et al. [15] investigated hardware implementations of 8X8 DCT and IDCT on different FPGA technologies using modified Loeffler algorithm. They simulated and synthesized the VHDL code with different FPGA families such as Xilinx, Altera and Lucent. Synthesis results for 8-point DCT/IDCT implementations were compared with SIF and 100 HDTV frames for all three FPGA families. Their implementation indicates significant speed of DCT based compression algorithms up to frames above the requirements of SIF, CCIR-TV, and HDTV frame formats.

\textbf{Santos} et al. [10] presented an FPGA implementation of a novel adaptive and predictive algorithm for lossy and hyper spectral image compression. They obtained an FPGA implementation of the lossy compression algorithm directly from a source code written in C language using CatapultC. CatapultC is a high level synthesis tool (HLS). They showed that how well the lossy compression algorithm performs on an FPGA in terms of throughput and area. Their results on a Virtex 4 4VLX200 show less memory requirements and higher frequency for the LCE algorithm.

\textbf{Sanjeev} et al. [8] emphasized on reducing the MSE (Mean Squared Error) and improving PSNR (Peak Signal-to-Noise Ratio) after the image compression. They surveyed on the image compression technique to find out the similarities and differences among them. They used the similar DCT transformation on images with little discussion on hardware implementation.

\textbf{Atitallah} et al. [16] compared the modified Loeffler algorithm and Distributed Arithmatic for implementing the DCT/IDCT algorithm for MPEG or H.26x video compression using VHDL. They implemented the design on Altera Stratix FPGA . They found that better results can be obtained with the modified Loeffler algorithm by using DSP blocks for the DCT/IDCT hardware implementation. 

\textbf{Roger} et al. [17] explored different architecture for DCT image compression using various adders. Their objective was to increase the JPEG compressor performance. They used carry lookahead, hierarchical carry lookahead and carry select architectures. The 2-D DCT architecture was synthesized in Altera FPGA. They found that the highest performance obtained for the 2-D DCT was 23\% higher than the original, using 11\% more logic cells.

% IMPLEMENTATION & EVALUATION
%=============================
\section{Implementation and Evaluation}

In this section we describe our implementation detail for each steps on image compression and decompression. Compression starts with the preprocessing of the image which we call color space conversion. Then DCT compression, quantization and encoding is done to get compressed image. This process is reversed to get the original image back. In this experiment, we have used Xilinx ISE web pack software for VHDL implementation and MATLAB [5] for image preprocessing.

%--------------------------------%--------------------------------%--------------------------------%--------------------------------%--------------------------------%--------------------------------

\subsection {Compression}

\subsubsection{Color Space Conversion}

We considered several images from the image folder which comes installed with an Windows computer and converted them into numerical values using Matlab. The process works same for both color and gray scale images. This preprocessing step is called color space conversion. The output after preprocessing is one matrix full of all the pixels from the original image. This converted matrix is saved in a text file so that it can be further put as input for the second step. In this step, we get the value of the matrix $P$.\\

%--------------------------------%--------------------------------
\subsubsection{2D DCT Implementation}

DCT is applied on the values obtained from the image transformation from the previous step. DCT converts the image pixel values from spatial region to frequency region. Initially, this frequency region is divided into several chunks of units. Each chunk consists of $8 X 8$ pixels. So, DCT compression algorithm will be applied on each chunk of the whole frequency domain iteratively. The DCT equation is the summation of the input function and cosine functions over $8 X 8$ block that is being compressed.

\begin{equation}
  \begin{split}
  DCT(\mathbf{u,v} )
    &= \frac{1}{4}C(u)C(v) \sum_{x=0}^{7}\sum_{y=0}^{7} P(x,y) \\
    &  \quad \cdot cos \Bigg[\frac{(2x+1)u\pi}{16}\Bigg] cos \Bigg[\frac{(2x+1)v\pi}{16}\Bigg]
  \end{split}
\end{equation}

Implementing this summation function can be approximated as:

\begin{equation}
  \begin{split}
  DCT(\mathbf{u,v} )
    &= C \cdot P \cdot {C}^T
  \end{split}
\end{equation}

Where,

\begin{equation}
 C(\mathbf{u,v} ) = \left\{
  \begin{array}{lr}
    \sqrt {\frac {2}{N}} & \quad \text{,  $u = 1$}\\
    \sqrt {\frac {2}{N}} cos \Bigg[\frac{(u-1)(2v-1)\pi}{2N}\Bigg] & \quad \text{,  $u \neq 1$}
  \end{array}
\right.
\end{equation}

$C$ is a constant matrix of the values of cosine obtained from the previous equation. Thus our implementation would consist of a matrix multiplication system that can multiply three matrices. $P$ refers to the input matrix of the image pixel values.\\

and

\begin{equation}
  \begin{split}
  {C}^T(\mathbf{u,v} )
    &= \sqrt {\frac {2}{N}} cos \Bigg[\frac{(2u-1)(v-1)\pi}{2N}\Bigg]
  \end{split}
\end{equation}

%--------------------------------%--------------------------------
\subsubsection{Quantization}

The quantization implementation requires us to implement division in VHDL. Quantization is defined as:

\begin{equation}
  \begin{split}
  {Q}_{DCT}
    &= round \Bigg(\frac{DCT(u,v)}{Q(u,v)}\Bigg)
  \end{split}
\end{equation}

The division function in the previous equation is not a matrix division but a scalar division. In a scalar division we only need to divide each number in the DCT matrix by the corresponding number in the quantization matrix. The quantization step is where all the data is lost. Depending on how much data loss is acceptable, the quantization matrix can be adjusted. Thus, the quantization matrix allows the user to retune the amount of compression required.\\

\begin{equation}
Q_{n} = \left\{
  \begin{array}{lr}
    \frac{100-n}{50}*Q_{50} & \quad \text{,  $n \ge 50$}\\
    \frac{50}{n} * Q_{50}& \quad \text{,  $n < 50$}
  \end{array}
\right.
\end{equation}

Where $n$ is the quantization level.

%--------------------------------%--------------------------------
\subsubsection{Entropy Encoding}

As entropy encoding is basically a zig-zag traversal of the 8x8 quantized blocks of values, it can be implemented as an address translation unit in hardware. The address translation unit is used when the values coming out of the Quantization stage of the pipeline are input into the registers of the Encoding stage of the pipeline.

%\begin{figure}[h!]
%  \begin{center}        
%        \includegraphics[width=0.4\textwidth]{zigzag}   
%	 \caption{The zig zag pattern for entropy encoding}        
% \end{center} 
%\end{figure}

Figure-3 shows how the matrix is transformed into linear values after zig-zag traversal over the whole matrix.\\

\subsubsection{Run Length Encoding}

Run Length Encoding is the point where the data compression really occurs. In the Run Length Encoding stage the image data arriving as a result of the previous stages is actually stored using a smaller number of bits. The out put from the Zig Zag encoding is converted into $1 \times 64$ standard logic vector that contains the occurrences of $0$'s and other values. Then the  number of 0's are  counted only from the vector and the other values are kept as it is which in turn provides significant savings in memory. Below is an example of RLE encoded example\\

%The input to this stage is a stream of numbers and the output from this stage is also a stream of numbers with the $(N+1)$ th value in the output stream being a number occurring in the input stream and the $N$ th value in the output stream being the number of continuous occurrences of that number. 

$Input: 4 \ 0 \ 0 \ 0 \ 9 \  0 \  0 \  0 \ 0 \  1\  1 \  0\  0 \  7 \  5\  0 \  0 \  0 \  0 \  0 \  0 \ 0 \  32  $ \

$Output: 4 \ 0 \ 3 \ 9 \ 0 \ 4 \ 1  \ 1 \ 0  \ 2 \ 7 \ 5 \ 0 \ 7 \ 32 $\\

In the hardware implementation used in this project this is implemented using an input and an output matrix. The input matrix is the zig-zag sequence of the values after they have been quantized and the output matrix is the run length encoding of the values in the input matrix.
%, with zeroes padded at the end of the output matrix.\

% Decompression Steps
%--------------------------------%--------------------------------%--------------------------------%--------------------------------%--------------------------------%--------------------------------

\subsection{Decompression}
Decompression steps are the opposite of the compression steps. We are providing little detail with example in the following sections.

%--------------------------------%--------------------------------
\subsubsection{Run Length and Entropy Decoding}
Run length decoding is done on the encoded file from the compressed image. The compact input data is spread as per the frequency of the zero's . This is the reverse process for the run length encoding step. For example, the below input is spread over in the output from this step.\\

$Input: 4 \ 0 \ 3 \ 9 \ 0 \ 4 \ 1  \ 1 \ 0  \ 2 \ 7 \ 5 \ 0 \ 7 \ 32 $\

$Output: 4 \ 0 \ 0 \ 0 \ 9 \  0 \  0 \  0 \ 0 \  1\  1 \  0\  0 \  7 \  5\  0 \  0 \  0 \  0 \  0 \  0 \ 0 \  32  $ \\

%$Input: 4 \ 5 \ 3 \ 0 \ 2 \ 1 $\
%
%$Output:  5 \ 5 \ 5 \ 5 \ 0 \ 0 \ 0 \ 1 \ 1$\\

This output becomes linear and need to be constructed as matrix format with $8 X 8$ blocks each. It is done by Zig Zag process. Each $64$ inputs are considered for reproducing the $8 X 8$ matrix. This is needed for the de quantization step to work on.

\subsubsection{Dequantization}

This process takes the input from decoded run length. Run length decoding and zig-zag process produces the initial quantized matrix. This matrix is again multiplied by the constant $Q_{75}$ matrix to get the inverse DCT coefficient matrix. The quantization matrix used in our experiment is given below.

 \begin{equation}
Q_{75} = \begin{bmatrix}
       20 &  1 &  7 &  9 & 4 & 2 & 0 & 2 \\
       4 &  13 &  -3 &  0 & 1 & 1 & 0 & 0 \\
       -19 &  -6 &  2 &  -9 & -3 & -1 & 0 & 0 \\
       -9 &  -14 &  -1 &  -1 & 2 & 1 & -1 & 0 \\
       2 &  4 &  -3 & -1  & 0 & 0 & 0 & 0 \\
       -1 &  4 &  -1 &  0 & 0 & 1 & 1 & 1 \\
       2 &  0 &  0 &  1 & 0 & 0 & 0 & 1 \\
       -2 &  0 &  0 &  -2 & 1 & 0 & 0 & 0\\
    \end{bmatrix}
\end{equation}

\subsubsection{Inverse DCT Implementation}
The quantized matrix is rounded up and multiplied by the constant matrix $C$ and $C^T$ to get the original pixel values. This step is called inverse DCT calculation. Inverse DCT equation is given below.

\begin{equation}
  \begin{split}
 P(\mathbf{x,y} )
    &= \frac{1}{4}\sum_{x=0}^{7}C(u)\sum_{y=0}^{7}C(v)  DCT(\mathbf{u,v} )   \\
    &  \quad \cdot cos \Bigg[\frac{(2x+1)u\pi}{16}\Bigg] cos \Bigg[\frac{(2x+1)v\pi}{16}\Bigg]
  \end{split}
\end{equation}

It can be simplified to,

\begin{equation}
  \begin{split}
 P(\mathbf{x,y} )
    &= round ( C^{-1} * IDCT (u,v) * (C^T)^{-1} ) + 128\\
  \end{split}
\end{equation}

Then, $128$ is added to each element of that result which gives the decompressed data file of the matrix $P(x,y)$ of the original $8 X 8$ image block. Finally, this matrix is processed in MATLAB to get the JPEG version of the image.

%\section{Controller and Datapath}
%
%\subsection{Datapath}
%
%Datapath diagram and description goes here. Datapath diagram and description goes here. 
%
%\subsection{Controller}
%
%Controller diagram and description goes here.
\section{Dataflow Diagrams}
\begin{figure*}%
\centering
\begin{subfigure}{.33\textwidth}
\centering
\includegraphics[height=2.7in]{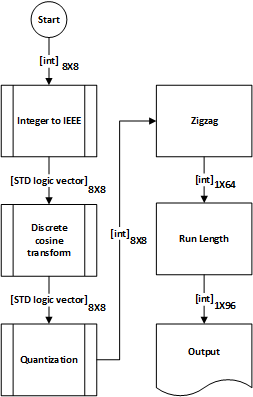}%
\caption{Complete process of image compression steps.}%
\label{compression_IEEE_1}%
\end{subfigure}\hfill%
\begin{subfigure}{.33\textwidth}
\centering
\includegraphics[height=2.7in]{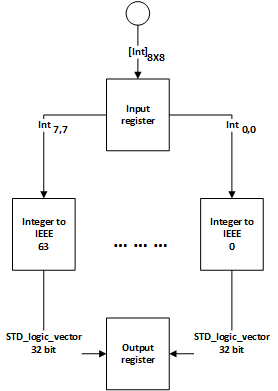}%
\caption{Integer to IEEE formatting.}%
\label{compression_IEEE_2}%
\end{subfigure}\hfill%
\begin{subfigure}{.33\textwidth}
\centering
\includegraphics[height=2.7in]{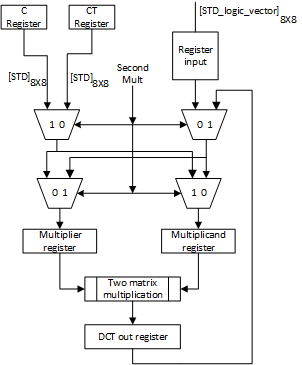}%
\caption{Discrete cosine transformation.}%
\label{compression_IEEE_3}%
\end{subfigure}%
\caption{Image compression, image foramtting, and DCT compression processes.}
\label{compression_IEEE}
\end{figure*}

This section describes the detailed data flow diagrams for the whole process. Figure~\ref{compression_IEEE_1} shows image compression process in detail. The formatting of image values to IEEE is done as explained in Figure~\ref{compression_IEEE_2}. The IEEE formatter provides standard logic vector output of 32 bits. Discrete cosine transformation is implemented as demonstrated in Figure~\ref{compression_IEEE_3}. 

\begin{figure*}%
\centering
\begin{subfigure}{.33\textwidth}
\includegraphics[width=\textwidth]{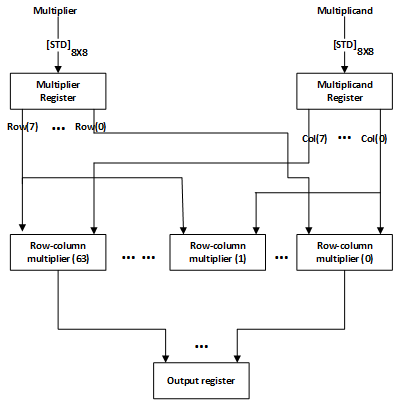}%
\caption{Two-matrix multiplication.}%
\label{matmul_1}%
\end{subfigure}\hfill%
\begin{subfigure}{.31\textwidth}
\centering
\includegraphics[height=2.4in]{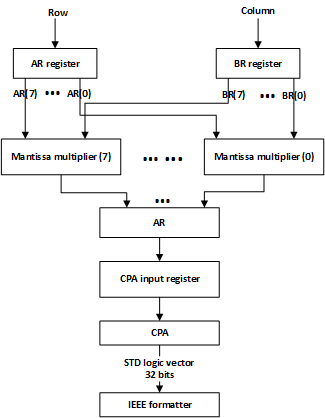}%
\caption{Row-column multiplication.}%
\label{matmul_2}%
\end{subfigure}\hfill%
\begin{subfigure}{.35\textwidth}
\includegraphics[height=2.4in]{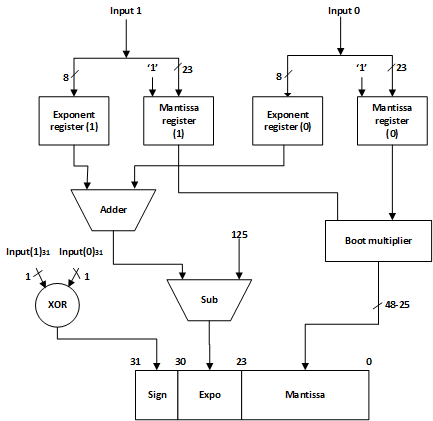}%
\caption{Mantissa multiplication.}%
\label{matmul_3}%
\end{subfigure}%
\caption{Matrix multiplication process.}
\label{matmul}
\end{figure*}

Discrete cosine transformation has a subprocess called two-matrix multiplication. This subprocess is implemented as shown in Figure~\ref{matmul_1}. Two-matrix multiplication subprocess contains 64 different row-column multiplier. Additionally, there are one multiplier and one multiplicand register that provide inputs for those 64 registers. From each row-column multiplier, 2 inputs are provided to total of 8 mantissa multiplier as explained in Figure~\ref{matmul_2}. Finally, mantissa multiplier provides 32 bit output which contains single sign bit, 7 exponent bits, and 24 mantissa bits as demonstrated in Figure~\ref{matmul_3}.

%--------------------------------%--------------------------------%--------------------------------%--------------------------------%--------------------------------%--------------------------------

\section{Synthesis and Simulation Results}

\begin{figure}[!h]
  \begin{center}        
        \includegraphics[width=0.48\textwidth]{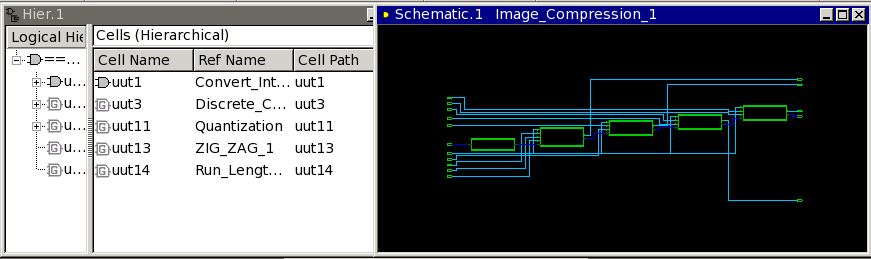}   
	 \caption{RTL schematic design of JPEG compression steps.}        
	 \label{rtlschem}
 \end{center} 
\end{figure}

We have used Synopsys Design Compiler for the logic synthesis. Additionally, we have parallelized the  implementation to decrease the execution time for speedup. The most significant factor for image compression is minimizing the processing time which is achieved in  our experiment. Although the area is increased due to this parallelism. Figure~\ref{rtlschem} shows the RTL schematic design of JPEG compression steps. The proposed implementation for DCT requires an area of 19064344.4 $\mu$ $m^2$ as shown in Tabel \ref{perf1}. In addition, the execution time for DCT implementation is 3.94 ns.

\begin{table}[!h]
  \centering
  \caption {Implementation and performance results.}
   \begin{tabular}{ ||>{\centering}p{2.5cm}|>{\centering}p{2.5cm}||}
    \hline 
    Parameters & Values \\ \hline
    Circuits &  DCT 2-D \\ \hline
    Area ($\mu$ $m^2$) &  19064344.4 \\ \hline
    Time  (ns) &  3.94 \\ \hline
    \end{tabular}
    
    \label{perf1}
\end{table}

%We will compare the compression performance with the standard image compression algorithms. We will compare our results with some contemporary works in this area as well.

  \begin{table}[!h]
   \caption {Image Compression Performance}
   \begin{tabular}{ || >{\centering}p{1.3cm} | >{\centering}p{1.2cm} | >{\centering}p{1.2cm} |  >{\centering}p{1.3cm} | >{\centering}p{1.1cm} ||} 
    \hline 
     Image (jpeg)   &   Dimension &  Original \ Size (KB) &  Compressed \ Size (KB) &  Reduction \ (Percent) \\ \hline 
    Desert  & 1024*768      &    846             &    127               &    84.98\% \\ \hline
    Koala      & 1024*768      & 781                & 160                  &    79.51\%     \\ \hline
    Lighthouse & 1024*768      & 561                & 100              &    82.17\%    \\ \hline 
    Penguins   & 1024*768      & 778                & 119               &    84.70\%    \\ \hline
    Tulips     & 1024*768      & 621                & 96                      &    84.54\%     \\ \hline
    \end{tabular}
    \label{perf2}
\end{table}

\begin{figure}[h!]
%\centering
\begin{subfigure}{.48\textwidth}
\includegraphics[width=\textwidth]{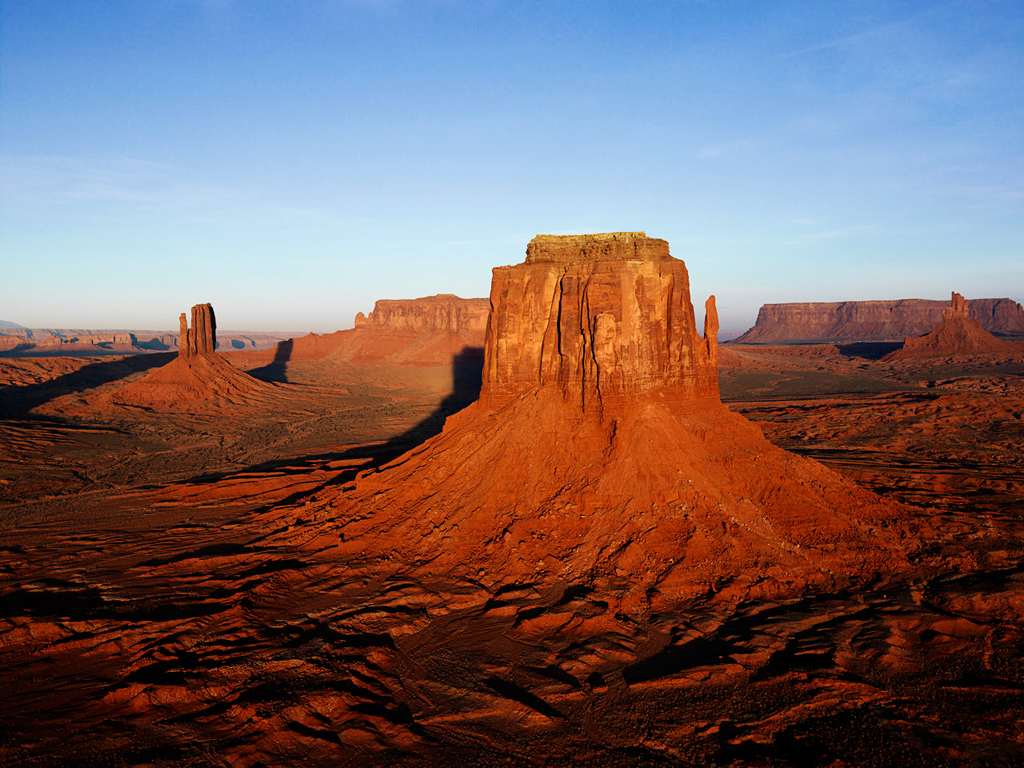}%
\caption{Desert (Before Compression)}%
\label{subfiga3}%
\end{subfigure}\hfill%
\begin{subfigure}{.48\textwidth}
\includegraphics[width=\textwidth]{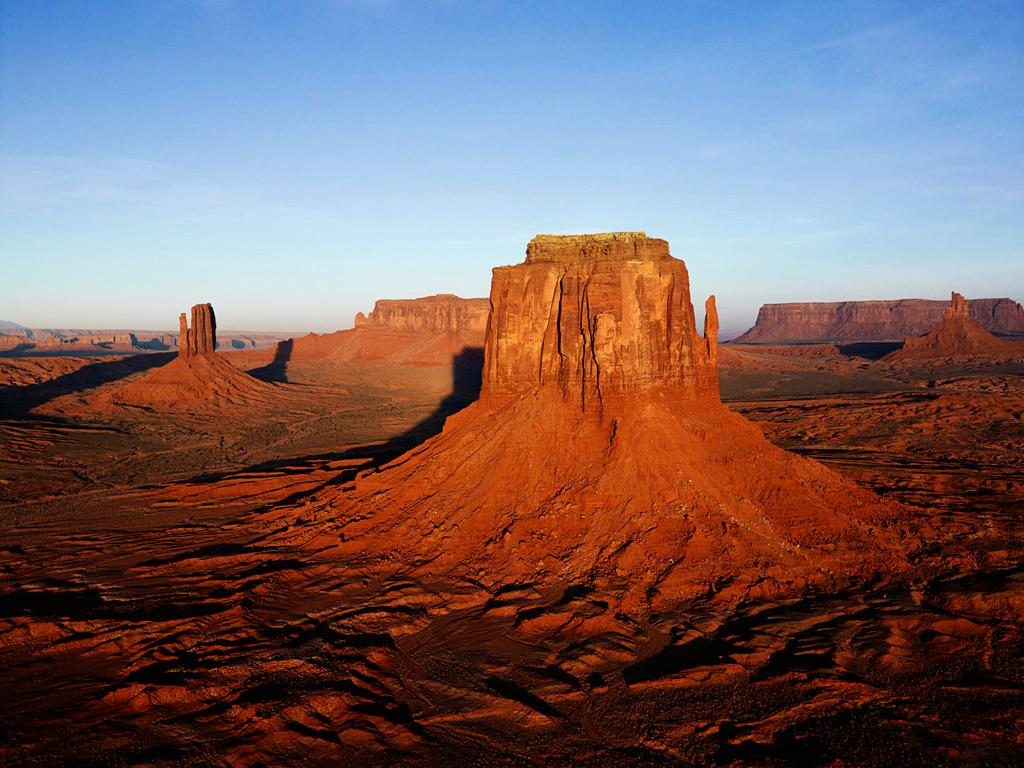}%
\caption{Desert (After Compression)}%
\label{subfigb}%
\end{subfigure}\hfill%
\caption{This image is collected from the windows images folder. (a) Input image before compression. (b) After image compression.}
\label{figabc12}
\end{figure}

Table~\ref{perf2} compares the size of JPEG image before and after compression and shows the performance of the algorithm with different image sizes, and with the quality of image after compression. Figure~\ref{figabc12} shows a sample image before and after compression.

\section{Conclusion}

DCT algorithm was simulated and synthesized in hardware aspect. With the definition of 2D-DCT, an approach for hardware implementation was developed. Then, a pure parallel structure for hardware realization was designed and approaches for accelerating multiplication and summation were proposed. Final results showed optimal hardware resource utilization and performance enhancement.

\appendices
%\section{}
%%
%%The project files including VHDL codes, sample images (both original and compressed) and data files for the matrices will be uploaded in the $github$ repository to help other researchers working in the same field.
%
%% you can choose not to have a title for an appendix
%% if you want by leaving the argument blank
%
%% APPENDIX 2
%
%
%
%\section{}
%
%We present original and compressed version of an sample image used in our experiment. Both the images are 1024 by 768 pixels by scaled down to fit in the width.

\section*{Acknowledgment}

The authors would like to thank the department of Computer Engineering, King Fahd University of Petroleum and Minerals, Saudi Arabia.

\ifCLASSOPTIONcaptionsoff
  \newpage
\fi

% References Section: Change references according to your need.
%----------------------------------------------------------------------------------------

\newpage

\end{document}